\documentclass[fleqn,12pt,twoside]{article}
\usepackage[headings]{espcrc1}

\usepackage{graphicx}

\newcommand{\jpsin}{$J/\psi$}
\newcommand{\psipn}{$\psi^\prime$}
\newcommand{\jpsi}{$J/\psi$ }
\newcommand{\psip}{$\psi^\prime$ }
\newcommand{\be}{\begin{equation}}
\newcommand{\ee}{\end{equation}}
\newcommand{\beq}{\begin{eqnarray}}
\newcommand{\eeq}{\end{eqnarray}}

\title{\jpsi Suppression}

\author{Marzia Nardi\address{Enrico Fermi Centre,
    Compendio Viminale,    00184 Roma (Italy)}\address{INFN, Via P. Giuria
    1, 10125 Torino (Italy)}
  \thanks{Present address: CERN, Phys. Dep, TH unit, 1211 Gen\`eve 23,
         Switzerland}
} 

\begin{document}

\maketitle

\begin{abstract}
I discuss the current theoretical interpretations of the anomalous
\jpsi suppression and their comparison to the most recent experimental data.
\end{abstract}

\section{Introduction}
\label{intro}

The behavior of $J/\psi$ mesons in a hot strongly interacting medium
was proposed as a test for its confinement status
\cite{Matsui:1986dk}: it was argued that the $J/\psi$, due to its
small size and strong binding energy, can not break up as a
consequence of interactions with normal hadrons, while in a deconfined
medium, the color screening dissolves the $c\bar{c}$ bond.

After this proposal, the study of the charmonium suppression in heavy
ion collisions has aroused a lot of interest.
With a careful and extensive analysis of the experimental data for
different interacting systems (from proton-proton to proton-nucleus
and nucleus-nucleus) it has been possible to observe a {\em normal}
suppression of the $J/\psi$ meson, presumably due to the absorption of
the preresonant $c\bar{c}$ state in the nuclear medium, in all
interactions up to S-U collisions and peripheral Pb-Pb. 
The experimental data, for proton-nucleus (p-A) collisions,
 can be accurately described by a simple
probabilistic formula for the \jpsi survival probability~\cite{KLNS}:
\be
S_{pA} \equiv \frac{\sigma^\psi_{pA}}{A\sigma^\psi_{pN}} =
\int d^2b \int_{-\infty}^{\infty} dz\,\rho_A(b,z)
\exp\left\{-(A-1) \int_z^{\infty}dz^\prime \,\rho_A(b,z^\prime)
\sigma_{abs}\right\}\, ,
\label{eq:survpA}
\ee
where $\sigma_{abs}$ is the effective cross-section for the absorption
of the \jpsi in nuclear matter: the experimental data fit gives
$\sigma_{abs}=4.18\pm 0.35$~mb~\cite{Alessandro:2004ap}; 
$\rho_A$ is the nuclear profile function, with parameters
tabulated in \cite{DeJager}.

The obvious generalization of Eq.~\ref{eq:survpA} to the
nucleus-nucleus (A-B) case~\cite{KLNS} gives an excellent description of
the \jpsi suppression in S-U as well as in peripheral Pb-Pb collisions.

In central Pb-Pb collisions, i.e. with impact parameter $b <
 8-8.5$~fm,
 a stronger suppression is
observed~\cite{Alessandro:2004ap}: this {\em anomalous} suppression is the
candidate as a signal of deconfined matter production at SPS.
Recent NA60 data in In-In collisions confirmed the departure from the
normal absorption in central collisions, with a pattern very similar
 to the one observed in Pb-Pb\cite{NA60}.
Surprisingly, also the first data from PHENIX (RHIC) show a
suppression of the \jpsi meson comparable with what observed at SPS
energies\cite{PHENIX,Drapier}, while the common expectation pointed to a much
stronger one.

The aim of this contribution is to give a critical overview of different
theoretical interpretations of the anomalous suppression, comparing them to
the most recent experimental data. In Section \ref{sec:DPM}, I will
present a popular model which explains the NA38/NA50 data without
assuming the deconfinement transition. Sections \ref{sec:therm} and 
\ref{sec:perc} present two different models of \jpsi suppression in a deconfined
medium: thermal dissociation and parton percolation. In Section
\ref{sec:regen}, I will discuss the regeneration model:
a recent approach which assumes that
the observed {\jpsin}'s can be produced not only in the initial
interactions but also in a subsequent stage of the system evolution.
This second contribution is expected to be marginal at SPS,
but very important at RHIC and LHC energies.
Finally, in Section \ref{sec:concl}, I will draw some conclusions of
what we learned in these first ten years of \jpsi suppression and what
we need from future experiments.

\section{General remarks}

It is known from proton-nucleon and pion-nucleon \cite{Antoniazzi:1992af}
(and, more recently, from electron-proton \cite{Abt:2002vq}) 
interactions that a large fraction (about 30\%-40\%)
 of the observed $J/\psi$'s are the decay products of higher
excited states of $c\bar{c}$ pairs ($\psi^\prime$, $\chi$).
Since the life-time of these quarkonium states is much larger than the
typical life-time of the medi\-um which is produced in the nucleus-nucleus
collisions, they decay in the vacuum.
Therefore this medi\-um 
(either hadronic gas or quark-gluon plasma)
sees not only the ground state
quarkonium, but also the different excited states, which have different
properties (size, binding energies) and different behavior: 
a smaller binding energy (and, consequently, a larger radius)
requires a lower dissolution temperature, in the case of a deconfined
medium; on the other hand, for the case of a hadronic system, a weakly
bound quarkonium state has a large break-up cross-section for
interactions with the other particles.

Therefore the final $J/\psi$ survival probability, to be compared to
the experimental data, has to be calculated as an average over the 
different components, each of them weighted with the corresponding
fraction $f_i$ of contribution to the observed $J/\psi$'s in the final state:

\be
S_{J/\psi}= f_{J/\psi} S_{J/\psi}^{dir}+f_{\chi}S_{\chi}^{dir}
+f_{\psi^\prime} S_{\psi^\prime}^{dir}
\label{eq:surv}
\ee

This fact has a very important consequence on the pattern of the 
$J/\psi$ suppression as a function of the centrality and of the energy
collisions and it must be considered for a careful comparison to
experimental data.

\section{Suppression by hadronic interactions}
\label{sec:DPM}

Before concluding that the anomalous suppression indicates the
formation of deconfined matter, one should check that the same affect
can not be reproduced in a normal hadronic scenario.

Many models of comover suppression have been proposed in the last
years, the crucial parameter is the $J/\psi$-hadron inelastic
cross-section $\sigma_{\psi h}$. The current theoretical estimates
are quite vague, ranging from 0.1 to a few mb !

Since in S-U interactions the observed suppression is compatible with
the normal one, the room for an additional suppression, due to
inelastic  interactions with secondary hadrons, is very small~:
this favors a very small $\sigma_{\psi h}$, since a large cross
section would have sizable effects.

On the other hand, the anomalous suppression in central Pb-Pb is very
strong and sets in quite abruptly~: to obtain this effect a comover
model must implement a mechanism which gives a very strong increase of
the comover density in going from S-U to Pb-Pb\footnote{One should, of
  course, make sure that the comover density needed to reproduce the
  data is not unrealistically large: this problem, unfortunately, is usually
  neglected. The authors of ref.~\cite{Maiani:2004qj} use a different
  approach: they set a physically meaningful limit for the density and
  the temperature of the hadron gas and conclude that it is not
  possible to describe the data of the anomalous suppression in the
  hadronic scenario}. 
The Dual-Parton Model (DPM)~\cite{DPM} is the
most effective one in this respect. 
The number of secondary hadrons is  the sum of two
contributions, proportional to the number of participating nucleons
and to the number of individual nucleon-nucleon collisions respectively:
\be
N_{com}(b,s,y,\sqrt{s})=C_1(y,\sqrt{s})N_{part}(b,s,\sqrt{s})+
C_2(y,\sqrt{s})N_{coll}(b,s,\sqrt{s}).
\label{eq:DPM}
\ee
In the previous formula, $b$ is the impact parameter, $s$ is the
2-dimensional coordinate in the transverse plane, $y$ is the rapidity
and $\sqrt{s}$ is the incident energy in the center of mass system.
The coefficients $C_1$ and $C_2$ are predicted in the DPM, as well as
their rapidity and energy dependence. 
The inelastic cross-section $\sigma_{\psi h}$ is fitted to the experimental data and
found to be 0.65 mb~\cite{Capella03}.
Predictions for RHIC energies are, of course, possible~\cite{Capella05}, 
with a correction taking into account
shadowing effects (needed to reproduce the data on hadron multiplicity).

In ref. \cite{Capella03} a reasonable description of S-U and Pb-Pb
data is shown. However the model fails in describing the In-In results
obtained by the NA60 Collab.~\cite{NA60}: the suppression predicted by
the DPM is much stronger than what is observed, signaling that the
effect of the collision term  in Eq.\ref{eq:DPM} is too strong: one
could try to reduce it by refitting the parameters, but then the 
agreement with Pb-Pb data would be lost.
The same problem occurs at RHIC energies, where the experimental data
are largely underestimated by the theoretical curve~\cite{Capella05}.
In my opinion, these data clearly disfavor the comover interactions as
an explanation for the observed \jpsi suppression.

\section{Thermal dissociation}
\label{sec:therm}

In sufficiently hot deconfined matter, color screening  dissolves
the binding of the quark-antiquark pair, and a stronger binding
energy requires a higher  temperature to be  dissolved.
On a microscopic level, it was argued that only a hot medium provides 
sufficiently hard gluons to dissociate the quark-antiquark bound
state, and this again implies a dissociation hierarchy as function of
the binding energy.
Another possible mechanism is the decay into open
charm mesons due to in-medium modification of mesonic
masses\cite{Digal:2001bh}.

Implicit, in this approach, is the assumption that the medium probed by the
quarkonium state is in thermal equilibrium.
Lattice studies show that the transition between confined and
deconfinement medium occurs at the critical temperature
$T_c \simeq 150-200$ MeV.

The large values of the charm (and bottom) quark mass allows potential 
theory to describe quite accurately the quarkonium spectroscopy\cite{Bali:2000gf}.

The authors of ref. \cite{Digal:2001bh} use the heavy-quark  potential $V(T)$
calculated on the lattice\cite{Karsch:2000kv} to obtain the quarkonium 
binding energy $M_Q(T)$
(i.e. the mass) and radius $r_Q(T)$, as a function of the temperature,
by solving the Schr\"odinger equation. 
Below the critical temperature $T_c$ one finds that the masses $M_Q(T)$
of the $c\bar{c}$ bound states are a decreasing function of the
temperature: 
when the condition $M_Q(T) < V_\infty(T)$ for a given bound state
 is satisfied ($V_\infty(T)$ is the
asymptotic limit of the potential for $r\to\infty$) the dissociation
process $Q\bar{Q}\to Q\bar{q}+\bar{Q}q$ is energetically favored and
that bound state dissolves.

Above the critical temperature $T_c$ one compares the binding radius
$r_Q(T)$ to the screening radius $1/\mu(T)$ (the distance at which the
interquark interaction vanishes, according to the lattice results):
again it is natural to assume that the bound state dissolves when 
$r_Q(T) > 1/\mu(T)$.
With these ingredients the authors of ref. \cite{Digal:2001bh} find
that $\chi$ and {\psip} dissolve already below the critical
temperature, while the \jpsi dissolves slightly above $T_c$. As a
consequence, a characteristic sequential suppression pattern is found,
with three steps for the onset of the suppression of the {\psipn}, $\chi$ and 
directly produced {\jpsin} respectively. One has to assume a model to 
translate the temperature into some experimental variable related to
the  centrality of the collision, and to take into account fluctuations 
in the number of participants and collisions for a given impact
parameter: the final result will be a smooth curve with two evident
drops, one for the $\chi$ and the \psip suppression (if they occur for
peripheral collisions they can hardly be separated)
 and the second one for the \jpsin.

%\begin{figure}
%{\includegraphics{diss.eps}}
%\caption{The $J/\psi$ suppression pattern as function of the
%  temperature, assuming 57\% of directly produced $J/\psi$'s, 35\% and
%  8\% of feed-down from $\chi_c$'s and $\psi^\prime$'s.}
%\label{fig:6}
%\end{figure}

%%%%%%%%%%%%%%%%%%%%%%%%%%%%%%%%%%%%%%%%%%%%%%%%%%%%%%%%%%%%%
\section{Parton percolation}
\label{sec:perc}

Hadrons are made by partons. When two ore more hadrons 
overlap, their partons interact. In a normal hadron-hadron
interaction, the overlapping phase is so short in time that
the partons separate again into hadrons before reaching an equilibrium
condition.
On the other hand, in a nucleus-nucleus collision, the number of
interacting hadrons is so high that their partons can interact several
times, they therefore loose their ``identity'', so that they do not belong
anymore to a particular hadron but form a big cluster of deconfined
medium: the quark-gluon plasma (QGP). 

The percolation theory is a
mathematical tool which studies how simple objects form clusters;
there are applications of the percolation idea in many physical
problems and the deconfinement transition in strongly interacting
matter is one of them.
For instance, in ref. \cite{Armesto:1996kt}
hadrons interact by exchanging color strings. When many hadrons
interact simultaneously in a small space-time region, these strings
overlap and, when their density reaches a critical value, they percolate.
The model of hadron interaction based on color string exchange 
(below the percolation threshold) is able to reproduce many 
features of experimental data (see references in \cite{Armesto:1996kt}).
This model therefore interpolate nicely from interactions in a 
normal hadronic medium and deconfined matter.
A similar approach was followed in \cite{Ugoccioni:2001rr}: 
a model of string fusion and percolation is used to describe several
experimental observables, including $J/\psi$ suppression.

The work of references
\cite{Nardi:1998qb,Digal:2002bm,Digal:2003sg}
is essentially focused on $J/\psi$ suppression by parton percolation,
inspired by lattice results where it was shown that the deconfined
transition in SU(2) Gauge Theory can be described by percolation 
of Polyakov Loops\cite{Satz:2001zf}.

Parton percolation is an essential prerequisite for QGP formation.
It should be noted that thermal equilibrium is not required and this
is main difference with respect to the approach described in section
\ref{sec:therm}.

Cluster formation in percolation theory shows critical behavior~: the
cluster size diverges at the percolation onset determined by the
critical density $n_c$ of overlapping objects. In a finite system 
the percolation onset is defined as the point at which the grow of the
cluster is more rapid \cite{Nardi:1998qb}.

The above considerations apply to a nucleus-nucleus collision if one
makes the following assumptions~: the overlapping objects, forming
clusters, are colored partons and the two-dimensional space where
they are distributed is the transverse plane projection of the
overlapping region of the two incident nuclei. 
A parton cluster represents a region in which
color charges can move freely: if it extends over the entire space,
one has, by definition, color deconfinement.

In high energy nuclear collision there is one additional difficulty:
the partons are emitted by the interacting nucleons, therefore their
distribution is not uniform in the transverse plane, but rather
reflects the original distribution of the initial nucleons, with
a higher concentration in the center than near the surface of the
nucleus. It is possible that in a
given collision, only the most central region of the produced
medium is dense and hot enough to allow for the deconfinement transition.
In this situation a local definition of percolation, as the one used
in ref. \cite{Digal:2002bm}, is more appropriated since the $J/\psi$ meson is very small
and therefore it is sensitive to the properties of a small spatial
region. The percolation approach, therefore, has this evident advantage:
 it naturally allows the deconfined matter to be formed
only in a limited part of the produced medium.

Having specified the objects which form clusters, the partons,
one has to describe their distribution in space
to apply the percolation idea to nuclear collisions. 
Since these partons are emitted by
the incident nucleons as a consequence of the strong interactions
during the first stages of the nuclear collision,
it is reasonable to assume that their number 
and spatial distribution 
is determined by the participating nucleons  from which they
originate, as done in refs \cite{Digal:2002bm,Digal:2003sg}. 
Therefore the density of partons is given by the product of
the participating nucleon density $n_s(b,A)$ (which depends on
the nucleus $A$ and on the impact parameter of the collision) 
and the number of partons per nucleon $dN_q(x,Q^2)/dy$
(the parton distribution
function, known from deep inelastic scattering experiments).
The fraction $x$ of the nucleon momentum carried by the parton 
 is related to the incident energy $\sqrt{s}$  by
$x=(k_T/\sqrt{s})$ (at midrapidity); $k_T$ is the average transverse
momentum of the parton and it is inversely proportional to its
transverse size. 
With these ingredients, the authors of ref. \cite{Digal:2003sg} find
that the critical cluster density is reached, in Pb-Pb collisions at
SPS, at $b\simeq 8$~fm and the corresponding value of the average
transverse momentum of the partons is $Q_c\simeq 0.7 $~GeV. The scales
of the charmonium states $\chi_c$ and $\psi^\prime$, given by the
inverse of their radii calculated in potential theory, are about 0.6
and 0.5 GeV respectively: they are therefore dissociated at the onset
of percolation. On the other hand, directly produced $J/\psi$'s have
smaller radii, therefore the average transverse momentum of the
deconfined partons must be at least 0.9-1.0 GeV to resolve them, so
only a denser medium, produced in more central collisions, 
 can dissociate them.

The $J/\psi$ survival probability in nuclear collisions is then
obtained, from the above considerations, by assuming that 
about 40\% of $J/\psi$'s (those
coming from $\chi_c$ and $\psi^\prime$ decays) produced inside the
percolating cluster disappear at the onset of percolation;
those formed outside the cluster, i.e. near the surface,
are not affected. The
remaining 60\%  of $J/\psi$'s (the directly produced ones)
survive until a cluster of hard enough partons is produced ($b\simeq
3-4$ fm). For a realistic comparison to the experimental data one has,
of course, to take into account impact parameter fluctuations (see,
for instance,\cite{KLNS}).
The result is in good agreement with the experimental data provided by
NA50 Collab. 

The same model has be used to predict the survival probability as
function of the centrality in
In-In collisions at SPS. One finds, in this case, that the percolation
onset is reached in semi-central collisions ($N_{part} \simeq 140$),
but the threshold for the dissociation of directly produced $J/\psi$'s
is never reached. One therefore expects, in this model, to observe a
suppression pattern with only one step.
The recently presented NA60 data show that this is indeed the case:
the suppression pattern has one step only and the amount of the
decrease is in remarkable agreement with the prediction of the
percolation model, but the onset is wrong. It seems therefore that
this model implements an excellent geometrical description of the
deconfined phase (the amount of the suppression reflects the fraction
of space occupied by the deconfined system) but needs to be improved.

\section{Suppression and Regeneration}
\label{sec:regen}

In the previous models it was assumed that the \jpsi are produced
exclusively upon the first, hard interactions of the colliding
nucleons. Recently, the idea that the \jpsi could also be created
 at the hadronization transition, by combining the 
$c\bar{c}$ pairs present in the plasma~\cite{Thews:2001em} has been
 proposed. 
This phenomenon is possible if the \jpsi states survive above the
transition temperature, as suggested by lattice and potential
calculations~\cite{Digal:2001bh,lattice}.

The authors of
refs. \cite{Grandchamp:2002wp,Grandchamp:2002iy,Grandchamp:2003uw}
consider both mechanisms~: the primordial \jpsi yield is subjected to
nuclear absorption followed by dissociation in the QGP phase.
A thermal contribution from statistical recombination of $c$ and
$\bar{c}$ quarks at the hadronization transition is added and also 
the \jpsi break-up due to interactions in the hadron gas is included.
The agreement with the experimental data of \jpsi suppression in Pb-Pb
collisions at SPS is excellent.
The result also shows that the regeneration process gives a small
contribution at SPS energies, while it is the dominant one at RHIC in
central Au-Au collisions.

\section{Discussion and Conclusions}
\label{sec:concl}

The recently presented experimental data allows us to make a few
considerations. First, they show that comover models can not
describe simultaneously the S-U, Pb-Pb and In-In data at SPS energies.
Even at RHIC energy they fail because they predict a suppression much
stronger of what is observed. Actually, every model proposed so far
which does not include the statistical regeneration fails
in the comparison to PHENIX data.
It would be tempting, therefore, to conclude that RHIC data
show evidence of the regeneration mechanism at work. However, this
conclusion is, in my opinion, premature. First of all, it is difficult
to understand, even in the regeneration scenario, how can the \jpsi
suppression both in Au-Au and Cu-Cu at different energies
($\sqrt{s_{NN}}= 62$ and 200 GeV) be equal: perhaps a different
explanation should be invoked. 

A closer look to the newest NA50 and NA60 data shows that the \jpsi
suppression, and more precisely
 the ratio of the number of measured \jpsin's to the
expected ones (considering also the normal suppression) is always
larger than 0.6. It was only in the previous data analysis that a few
experimental points descended below this value for very central
collisions, suggesting the presence a `second drop' due to the
suppression of directly produced \jpsin's.

The newest SPS data show no evidence of  direct \jpsi suppression~:
the experimental data could be explained, with Eq.\ref{eq:surv}, by
assuming $S_{J/\psi}^{dir}=1$. This is confirmed by the agreement of
the result of the percolation model with the NA60 data in central
In-In collisions~: in this model only \psip and $\chi$ are suppressed, because the
\jpsi onset is out of reach in this system. The fact that the suppression in central
In-In is about 0.8 instead of 0.6 (as it should be if one removes
totally the \psip and $\chi$ components) is due to finite size
effects: the deconfinement sets in only in the central, hottest region
and only the \psip and $\chi$ formed inside are affected.

This observation is corroborated by recent lattice 
calculations\cite{lattice}, where
the dissociation temperature for the \jpsi in the QGP is found to be
very high,
about $1.6-2T_c$. Similar results have been found in recent potential
model calculations, as shown in refs. \cite{Wong:2004zr,Alberico:2005xw}.
The practical consequence of this fact is that the energy
density required to break up the directly produced \jpsin's 
can be as high as 30 GeV/fm$^3$, definitely out of reach at SPS and
even in central Au-Au collisions at RHIC ! If this is the case, we
should be able to see the suppression of the direct \jpsi component
only at LHC.

In this way it is easy to understand why the suppression observed at
RHIC is quantitatively similar to the one seen at SPS: since the \psip
and $\chi$ components are already totally removed (except for finite
size effects) at SPS, they can not be {\em more} suppressed at
RHIC. \jpsin's are unaffected both at SPS and at RHIC.
Obviously, more precise data from RHIC experiments are necessary to
confirm this scenario. 

The reason why all theoretical models not including regeneration
effects underestimate the data is simply that all of them assume
that the direct \jpsi component can be suppressed: partially at SPS
and more intensely at RHIC.

It seems, therefore, that RHIC data can be explained without invoking
the regeneration of \jpsin's. Of course, one should check this idea,
by looking, for instance, to other observables related to the \jpsin,
like flow, rapidity and $p_T$ distribution.
The regeneration model has already provided
quantitative predictions in this respect\cite{Rapp:2005rr,Thews:2005vj}. 

Finally, I would like to present a few comments about the
deconfinement models in general. It is very common to assume that all
charmonium states inside the QGP bubble are inexorably and 
instantaneously suppressed.
This is, of course, an extreme hypothesis: even in a deconfined phase,
some time is required to break up a \jpsi (or \psip or $\chi$) meson,
at least the time needed for the $c$ and $\bar{c}$ to fly apart from
each other. If the QGP lifetime is shorter of this time or if the
$c\bar{c}$ is produced near the QGP surface with a transverse momentum
enough to escape before being affected, 
then the corresponding charmonium state has a chance
to survive even in the deconfined medium~! These effects are
reasonably very important in small system (like In-In) or in
peripheral collisions, right at the onset of the deconfinement.
It was reasonable, so far, to neglect these corrections to keep under
control the number of free parameters of the theory, but the level of
accuracy reached now by the latest data forces us to seriously think 
about this necessary step forward in future developments.

\end{document}